\input{aipcheck}

\documentclass[
    ,final              ]{aipproc}

\layoutstyle{8x11double}

\begin{document}

\title{Distributional cosmological quantities solve the paradox of soft
singularity crossing}



\author{L\'{a}szl\'{o} \'{A}. Gergely}
{address={Departments of Theoretical and Experimental Physics, University
of Szeged, Tisza Lajos krt 84-86, Szeged 6720, Hungary} }

\author{Zolt\'{a}n Keresztes}
{address={Departments of Theoretical and Experimental Physics, University
of Szeged, Tisza Lajos krt 84-86, Szeged 6720, Hungary} }

\author{Alexander Yu. Kamenshchik}
{address={Dipartimento di Fisica and INFN, via Irnerio 46, 40126 Bologna,
Italy} ,altaddress={L. D. Landau Institute for Theoretical Physics, Russian
Academy of Sciences, Kosygin street 2, 119334 Moscow, Russia} }

\begin{abstract}
Both dark energy models and modified gravity theories could lead to
cosmological evolutions different from either the recollapse into a Big
Crunch or exponential de Sitter expansion. The newly arising singularities
may represent true endpoints of the evolution or alternatively they can
allow for the extension of geodesics through them. In the latter case only
the components of the Riemann tensor representing tidal forces diverge. A
subclass of these soft singularities, the Sudden Future Singularity (SFS)
occurs at finite time, finite scale factor and finite Hubble parameter, only
the deceleration parameter being divergent. In a Friedmann universe evolving
in the framework of general relativity they are realized by perfect fluids
with regular energy density and diverging pressure at the SFS. A particular
SFS, the Big Brake occurs when the energy density vanishes and the expansion
arrives at a full stop at the singularity. Such scenarios are generated by
either a particular scalar field (the tachyon field) or the anti-Chaplygin
gas. By adding any matter (in particular the simplest, the dust) to these
models, an unwanted feature appears: at the finite scale factor of the SFS
the matter energy density remains finite, implying (for a spatially flat
universe) a finite Hubble parameter, hence finite expansion rate, rather
then full stop. The universe would then further expand through the
singularity, this nevertheless seems forbidden as the energy density of the
tachyonic field / anti-Chaplygin gas would become ill-defined. This paradox
is relieved in the case of the anti-Chaplygin gas by redefining its energy
density and pressure in terms of distributions peaked on the singularity.
The regular cosmological quantities which are continuous across the SFS are
then the energy density and the square of the Hubble parameter; those
allowing for a jump at the SFS are the Hubble parameter and expansion rate
(both being mirror-symmetric). The pressure and the decelaration parameter
will contain Dirac delta-function contributions peaked on the SFS, however
this is no disadvantage as they anyhow diverge at the singularity.
\end{abstract}

\maketitle


\section{Introduction}

General relativity and the Copernican principle, combined with observations
on the Hubble redshift of the galaxies and modelling the present baryonic
content of the Universe by a pressureless perfect fluid (dust) together with
a minor contribution from radiation leads to a Universe born from a Big Bang
singularity at finite time in the past. The Big Bang is characterized by
infinite values of the energy density $\rho $, pressure $p$ and temperature $%
T$. The scale factor $a$, characterizing the size of the Universe vanishes,
leading to a diverging scalar curvature, a true singularity.

The study of the rotation curves of galaxies, the stability of galaxy
clusters and the formation of structure in the Universe all imply the
existence of a dark matter component, manifesting itself only through the
gravitational interaction and dominating over baryonic matter by
approximately a factor of ten. Dark matter can be either cold or warm (but
not hot), and its inclusion into the past cosmological evolution does not
eliminate the Big Bang singularity. It is also confirmed by the very
existence of the cosmic microwave background and light element abundancies
in the Universe.

The future can be either continued expansion (still persisting after
infinite time or just asymptoting to a stop) or an expansion arriving to a
halt after finite time, followed by a contraction phase, leading eventually
to a Big Crunch singularity, which is very similar to the Big Bang. The
actual scenario is selected by the amount of combined dark and baryonic
matter densities, as compared to the critical density. In all these
scenarios the future evolution is decelerated due to gravitational
attraction.

Modern cosmological observations (distant supernovae of type Ia, the cosmic
microwave background, gravitational lensing) confirm on one hand that the
Universe is quite close to the critical energy density ($k=0$), however they
imply the necessity of an accelerated expansion in the recent past \cite%
{cosmic}, eventually disruling the three above mentioned scenarios for the
future of the Universe. Introducing dark energy, accounted for $73\%$ of the
energy content of the Universe (dark and baryonic matter contributing with $%
23\%$ and $4\%$, respectively) leads to the observed accelerated expansion
as it violates the strong energy condition. The energy density $\rho $ and
the pressure $p$\ of the dark energy satisfies $\rho +3p<0$, the condition
for accelerated expansion imposed by the Raychaudhuri equation%
\begin{equation}
\frac{\ddot{a}}{a}=-\frac{1}{2}\left( \rho +3p\right) ~  \label{Ray}
\end{equation}%
(we chose units $c=1$ and $8\pi G/3=1$).

In its simplest variant dark energy is the cosmological constant $\Lambda $,
with negligible contribution to the dynamics of the Universe in the past,
however modifying its future. In the $\Lambda $CDM (cosmological constant
and cold dark matter) model the Universe asymptotes to an exponentially
expanding de Sitter universe. Although of appealing simplicity, a
cosmological constant would conflict by many orders of magnitude the outcome
of all variants of calculation of the vacuum expectation energy. A dynamic
dark energy model would be clearly more satisfactory and perfectly
compatible with observations (which settle but the present value of this
field). For a review on dark energy models see \cite{dark}. There are many
dark energy candidates, their common feature being that they change the
future of the Universe in a drastic manner.

In Section 2 we enlist and succinctly characterize the possible outcomes of
such dynamic dark energy dominated evolutions together with certain
unconventional evolutions in modified gravity theories, leading to various
exotic singularities. In Section 3 we concentrate on a particular type of
evolution, leading to a Big Brake singularity. By adding ordinary 
matter to the model, the Big Brake singularity is
generalized to a Sudden Future Singularity (SFS). This is still a soft and
traversable singularity, however the future evolution is
obstructed by the dark energy becoming ill-defined. A possible
way of overcoming this difficulty is by generalizing the cosmological
quantities in a distributional sense. 

\section{A compendium of exotic cosmological singularities}

The common characteristic of all dark energy induced, novel type of
singularities is that they occur in finite time. Despite certain components
of the Riemann curvature tensor diverging some of these singularities remain
traversable. The classification below based on traversability is consistent
with Kr\'{o}lak's definition of the strongness of a singularity \cite%
{Krolakstrong}.

\subsection{Strong singularities}

These are the singularities of type I. and III. in the classification of
Ref. \cite{classification}.

Singulatities of type I. occur for phantom dark energy models (with
barotropic index $w=p/\rho $ slightly smaller than $-1$). These models have
the counter-intuitive feature that the energy density increases with the
expansion of the Universe. The singularity, dubbed Big Rip or Doomsday \cite{doom}
occurs at finite time and infinite scale factor $a$ and is characterized by
diverging Hubble parameter $H=\dot{a}/a$ and a diverging $\dot{H}$. Due to
the Raychaudhuri equation (\ref{Ray}) and the Friedmann equation%
\begin{equation}
\left( \frac{\dot{a}}{a}\right) ^{2}=\rho  \label{Fried}
\end{equation}
then both$~\rho $ and $p$ also diverge. The energy density and pressure thus
behave similarly as in the Big Bang or Big Crunch, however this happens at
infinite, rather than vanishing scale factor.

Singularities of type III. are very similar, $H,~\dot{H},~\rho $ and $p$
diverge, however this occurs at finite scale factor. Therefore the
singularities of type III. are also known as Finite Scale Factor
singularities.\footnote{%
Nevertheless, a finite scale factor is also characteristic for other
singularities. In fact all weak (soft) singularities to be mentioned in this
paper occur not only at finite time, but also finite (but non-vanishing)
scale factor.} Note that although this singularity is strong according to Kr%
\'{o}lak's definition, it shows up as weak according to Tipler's definition 
\cite{Tiplerstrong}, which seems then less adequate to characterize the
strongness of a singularity. The singularities of type III. are compatible
with available cosmological observations \cite{Denkiewicz}.

A particular singularity of type III. is the Big Freeze, occurring in the
evolution of the generalized phantom Chaplygin gas \cite{BigFreezeDemar}.

\subsection{Weak singularities}

Pure kinematical investigations of evolutions in a Friedmann universe lead
to the possibility of Sudden Future Singularity (SFS) occurrence \cite%
{sudden}. Such singularities are of type II in the classification of Ref. 
\cite{classification} and are characterized by finite scale factor $a$ and
finite Hubble parameter $H$, while $\dot{H}$ diverges. Hence at these
singulatities the energy density is finite, while the pressure diverges. As
the metric contains only the scale factor, the geodesic equations will
contain but $H$, hence point particles may pass through this singularity,
generating afterwards the new geometry. The diverging $\dot{H}$ appears only
in the deviation equation, generating infinite tidal forces at the SFS
crossing, but only for an infinitesimally short time \cite{crossing}. SFS
are weak in both the Kr\'{o}lak and Tippler's definitions.

A particular SFS occurs when a full stop is realized at the singularity.
Such Big Brake singularities could be produced by the dynamics of an
anti-Chaplygin gas or by a particular tachyonic scalar field showing
superluminal evolution over certain periods of its existence \cite{tach0}.
The tachyonic model does not violate causality due to the continued
homogeneity and isotropy of the Universe and was shown to be in agreement
with observations on the supernovae of type Ia \cite{tach1}. The Big Brake
occurs after a time comparable with the present age of the Universe and was
also shown explicitly to be traversable and to eventually evolve into a Big
Crunch \cite{tach2}. In Refs. \cite{quantum} the solutions of the
Wheeler-DeWitt equation for the quantum state of the universe in the
presence of the Big Brake singularity was studied.

A time-reversed version of the Big Brake singularity is the Big D\'{e}%
marrage \cite{BigFreezeDemar}, when the Universe starts expanding from a
state of infinite pressure but finite energy density.

There are also weak singularities characterized by vanishing pressure and
vanishing energy density, however their ratio, the barotropic index $w$
being divergent$.$ Both the singularities of type IV. from the
classification of Ref. \cite{classification}, where the time derivatives of
rank three or higher of the scale factor diverge; and the w-singularities
with completely regular scale factor introduced in Ref. \cite{w}\ belong
here. These singularities are quite soft, they do not harm in any way the
evolution of the Universe or standard matter, rather manifest themselves
only in the dark energy model, possibly signaling its breakdown.

\subsection{Exotic brane-world singularities}

In brane-worlds the Einstein equation is replaced by the effective Einstein
equation. Beside the cosmological constant term and the energy-momentum
tensor this equation has additional source terms: i) a quadratic source in
the energy-momentum tensor (which becomes important only at high energy
densities or pressures), ii) a pull-back to the brane of non-standard model
fields acting in 5 dimensions, iii) the asymmetric embedding of the brane.
All these are reviewed in detail in \cite{Decomp}.

A singularity very similar to the SFS, dubbed quiescent singularity arises
in brane cosmology, in which $\rho $ and $H$ remain finite, but all higher
derivatives of the scale factor diverge as the cosmological singularity is
approached \cite{ShtanovSahni}.

Brane-world dynamics however, in particular the presence of the
energy-momentum squared term among the source terms allows for the
appearance of even stranger singularities, which are characterized by
diverging $\rho $ and $p$, nevertheless regular evolutions of the scale
factor. Such an example is provided by the collapse of a perfect fluid
metamorphosing into dark energy \cite{collapse}.

Another such singularity arises in the context of brane-world flat
Swiss-cheese cosmologies, in the presence of a huge cosmological constant 
\cite{BlackString}. At this singularity the scale factor, its first, second
and all higher derivatives stay regular. This universe forever expands and
decelerates, as its general relativistic analogue, the Einstein-Straus model 
\cite{EinsteinStraus}. However after a finite time the pressure diverges to
plus infinity. This smooth pressure singularity is different from the case
when both the pressure and the second derivative of the scale factor
diverge, the latter stays regular. The accompanying energy density turns
negative shortly before reaching the singularity and becomes ill-defined
there. The asymmetric embedding enhances the apparition of such a
singularity. There is a critical value of the asymmetry in the embedding,
above which these singularities necessarily appear \cite{AsymSwissCheese}.

If one combines the cosmological constant, the energy-momentum and the
energy-momentum squared source terms into an effective fluid, it turns out
that this is dust, following the standard evolution of an Einstein-Straus
model. The effective energy density evolves through positive values through
the singularity, towards reaching asymptotically zero, as the universe
expands. In terms of the effective dust source it is quite natural that the
singularity can be passed through. Nevertheless the pressure of the physical
fluid diverges and its energy density becomes ill-defined. The singularity
is induced by the brane dynamics non-linear in the energy-momentum, modified
as compared to GR.

\section{SFS crossing}

The Big Brake singularity is the simplest SFS and phenomenological models,
like a tachyonic scalar field or anti-Chaplygin gas were found, which evolve
into a Big Brake \cite{tach0}. Although the tachyonic scalar field has a
subluminal evolution at present and mimics well dark energy \cite{tach1},
also displays a dust-like (dark matter like) behaviour in the more distant
past \cite{tach2}, a more comprehensive cosmological model would certainly
include baryonic matter as well, customarily modelled by dust. The addition
of dust to the tachyonic scalar field however induces a paradox. Its energy
density at any finite scale factor being positive, by virtue of the
Friedmann equation the Hubble parameter will not vanish at the singularity.
The Big Brake is replaced by a SFS exhibiting a finite expansion rate. The
paradox arises from allowing for further expansion: for larger scale
factor than the one characterizing the SFS the tachyonic field becomes
ill-defined. The same paradox also arises when the dust is added to the
anti-Chaplygin gas.

In Ref. \cite{aCh}, based on certain distributional identities we have 
worked out the details of including a distributional contribution to 
the pressure of the anti-Chaplygin gas (and equivalently to $\dot{H}$), 
centered on the SFS: 
\begin{equation}
p_{ACh}=\sqrt{\frac{A}{6H_{S}|t_{SFS}-t|}}+\frac{4}{3}H_{S}\delta
(t_{SFS}-t)~,  \label{pressure-new}
\end{equation}%
\begin{equation}
\dot{H}=-2H_{SFS}\delta (t_{SFS}-t)-\sqrt{\frac{3A}{8H_{SFS}a_{SFS}^{4}}}%
\frac{sgn(t_{SFS}-t)}{\sqrt{|t_{SFS}-t|}}~.  \label{Hubble-der-sing}
\end{equation}%
Then \thinspace $H$ may have a jump (the derivative of the Heaviside
function being a delta function). In order to keep the energy density
continuous, $H^{2}$ should not have a jump, thus when crossing the SFS, the
Hubble parameter should obey a $Z_{2}$-symmetry. If the Universe arrives to
the SFS with the expansion rate $\dot{a}_{SFS}$, after crossing it it will
have the expansion rate $-\dot{a}_{SFS}$. The respective equations are:%
\begin{eqnarray}
H(t) &=&H_{SFS}sgn(t_{SFS}-t)  \nonumber \\
&&+\sqrt{\frac{3A}{2H_{SFS}a_{SFS}^{4}}}sgn(t_{SFS}-t)\sqrt{|t_{SFS}-t|}.
\label{Hubble-sing2}
\end{eqnarray}%
In order to preserve the anti-Chaplygin gas equation of state $p=A/\rho $ a
delta function also enters the denominator of the energy density.
Alternatively, $\rho $ may be kept regular, but then the equation of state
should be generalized into a distributional relation.

There is a full analogy with a ball bouncing back from a wall or a tennis /
squash racquet. A simple description of the process includes a sudden
reversal of the normal velocity. A detailed description instead requires to
allow for modelling the ball deformation. After being compressed, the ball
will reach a full stop, before bouncing back. A description of the SFS
crossing without distributions would require to deform the equation of state
in the 2--component fluid, such that $H=0$ occurs at the SFS.


\begin{theacknowledgments}
  L\'{A}G is grateful for the organizers of the MultiCosmoFun12 Conference 
for invitaton and financial support. We acknowledge useful discussions 
with V. Gorini, M. O. Katanaev, V. N. Lukash, U. Moschella, D. Polarski and 
A. A. Starobinsky. This work was supported or partially supported by 
European Union / European Social Fund grant 
T\'{A}MOP-4.2.2.A-11/1/KONV-2012-0060 (L\'{A}G),
OTKA grant no. 100216 (ZK), 
and the RFBR grant no. 11-02-00643 (AYK).

\end{theacknowledgments}


\end{document}